\documentclass[twocolumn,prd,superscriptaddress,preprintnumbers,nofootinbib]{revtex4-1}[11pt]
\usepackage{amsmath,amssymb,graphicx}
\graphicspath{{figs/}}
\usepackage{epsf,verbatim}
\usepackage{hyperref}
\usepackage{comment}
\usepackage{color}
\usepackage{slashed}
\usepackage{subfigure}
\usepackage{comment}
\usepackage{mdwlist, paralist}
\usepackage{rotating}
\usepackage{bm}
\usepackage{multirow}
\usepackage{lipsum}

\newcommand{\beq}{\begin{equation}}
\newcommand{\eeq}{\end{equation}}
\newcommand{\bal}{\begin{align}}
\newcommand{\eal}{\end{align}}
\newcommand{\bit}{\begin{itemize}}
\newcommand{\eit}{\end{itemize}}
\newcommand{\ben}{\begin{enumerate}}
\newcommand{\een}{\end{enumerate}}

\renewcommand{\eqref}[1]{Eq.~(\ref{#1})}

\newcommand{\abs}[1]{\left| #1 \right|}

\begin{document}     

\title{Gravothermal Phase Transition, Black Holes and Space Dimensionality}

\author{Wei-Xiang Feng}
\email{wfeng016@ucr.edu}
\affiliation{%
Department of Physics and Astronomy, University of California, Riverside, CA 92521, USA
}%



\begin{abstract}
In the framework of gravothermal evolution of an ideal monatomic fluid, 
I examine the dynamical instability of the fluid sphere in ($N$+1) dimensions by exploiting Chandrasekhar's criterion to each quasistatic equilibrium along the sequence of the evolution. Once the instability is triggered, it would probably collapse into a black hole if no other interaction halts the process.
From this viewpoint, the privilege of (3+1)-dimensional spacetime is manifest, as it is the \emph{marginal dimensionality} in which the ideal monatomic fluid \emph{is stable but not too stable.} 
Moreover, it is the \emph{unique} dimensionality that allows stable hydrostatic equilibrium with \emph{positive cosmological constant}.
While all higher dimensional ($N>3$) spheres are genuinely unstable. 
In contrast, in (2+1)-dimensional spacetime it is \emph{too stable} either in the context of Newton's theory of gravity or Einstein's general relativity. 
It is well known that the role of \emph{negative cosmological constant} is crucial to have the Ba{\~n}ados-Teitelboim-Zanelli (BTZ) black hole solution and the equilibrium configurations of a fluid disk. Owing to the negativeness of the cosmological constant, there is no unstable configuration for a homogeneous fluid disk to collapse into a naked singularity, which supports the cosmic censorship conjecture. However, BTZ holes of mass $\mathcal{M}_{\rm BTZ}>0$ could emerge from collapsing fluid disks.
The implications of spacetime dimensionality are briefly discussed.
\end{abstract}

\pacs{Valid PACS appear here}
                                               
\maketitle


\paragraph*{Introduction.---} 
Black holes (BHs) are the most profound prediction of Einstein's general relativity (GR), though their existence is questionable in the early days. 
Oppenheimer and Snyder~\cite{Oppenheimer:1939ue} first demonstrated the dynamical process of forming a BH from dust collapse, and the spacetime singularity is inevitable. However, the idealized collapsing process of spherical symmetry might be unrealistic. Until 1965, Penrose showed~\cite{Penrose:1964wq} mathematically that whenever matter satisfies reasonable energy conditions, a BH is a generic consequence of GR regardless of spherical symmetry.
In astrophysics, a BH is the end state of a conventional star running out of fuel. But it could also emerge from the direct collapse of clouds of gas without igniting a nuclear reaction.

In the context of Newtonian gravity (NG), the self-gravitating ideal monatomic fluid is too stable~\cite{Shapiro:1983du}. The pressure always counteracts gravitational attraction and stabilizes the fluid.
In GR, the pressure in the fluid is a double-edged sword.  Because not only the energy density but also the pressure is sourcing gravity, once the pressure starts to dominate energy density at some point, it will destabilize the fluid. This is why the instability and collapse into BHs from a fluid can happen in the framework of GR.
All of the above concern BHs in (3+1)-dimensional spacetime. Of course, there is nothing to hinder theorists from considering BHs in ($N$+1) dimensions with $N\neq3$. 
BHs in higher dimensions have been studied thoroughly in the literature~\cite{Myers:1986un,Emparan:2008eg,Horowitz:2012nnc}. 
Although the stability of a fluid sphere and its dimensional constraint has been explored somewhat~\cite{Rahaman:2014gaa}, less investigated is the instability condition of the higher-dimensional BHs coming from collapsing fluids. As we shall see, the ideal fluids in the context of NG are genuinely unstable for $N>3$, and the GR effect makes the situation worse. However, the presence of cosmological constant $\lambda$ will modify the situation, in particular, $\lambda<0$ can stabilize the fluid sphere.
We also note that the dynamical instability of stellar equilibrium for $N=3$ with cosmological constant was  studied to some extent in Refs.~\cite{Boehmer:2005kk,Bordbar:2015wva,Posada:2020svn}.

In lower dimensions, {i.e.}, $N=2$, gravity is bizarre. 
The Ba{\~n}ados-Teitelboim-Zanelli (BTZ) BH solution exists only if a ``negative'' cosmological constant $\lambda=-1/\ell^2<0$ is introduced, where $\ell$ is the background radius of curvature~\cite{Banados:1992wn}. This can be understood from  the unit of Newton's constant in ($N$+1) dimensions: $[G_N]=[{\rm M}]^{-1}[{\rm L}]^{N}[{\rm T}]^{-2}$. For $N=2$, setting $c=1$ determines the fundamental ``mass scale'' in terms of the Newton's constant $G_2$, but the fundamental ``length scale'' cannot be settled down. Thus an independent length scale $\ell=(-\lambda)^{-1/2}$ must be introduced independently. Moreover, the negativeness of $\lambda$ permits the BH solution~\cite{Ida:2000jh}.
In addition, there is no Newtonian limit in (2+1) dimensions. Gravity has no local degrees of freedom (locally flat), thus no gravitational wave (or graviton) can propagate. This reflects the fact that static particles do not gravitate~\cite{Giddings:1983es,Deser:1983tn,Deser:1983nh}.
In contrast, the collective behavior of thermal particles will gravitate and demand the fluid description under its self-gravity. Crucially, $\lambda<0$ is also to guarantee the hydrostatic equilibrium (the pressure is monotonically decreasing)~\cite{Cruz:1994ar}.

The basic mechanism of hierarchical structure formation (stars, galaxies, halos, etc.) relies on the Jeans instability~\cite{Jeans:1902fpv}, which determines the largest mass (Jeans mass, also called Bonnor-Ebert mass~\cite{Bonner:1956sv,Ebert:1955}) of an isothermal gas sphere can still remain in hydrostatic equilibrium. For the gas sphere heavier than this, it will further collapse or fragment into smaller and denser objects~\cite{Chavanis:2001hd}. Then it transitions into \emph{gravothermal evolution}. As a gravitationally bound system, it gets hotter and hotter as it releases thermal energy through dissipation~\cite{LyndenBell:1968yw,Hachisu:1978gravothermal,Spitzer:1987aa}. If the mass sphere is sufficiently heavy ($\gtrsim10^6~M_\odot$), this process will persist without triggering the thermonuclear sources of energy, while it behaves as a ``supermassive star.'' Nevertheless, the gravothermal evolution will end eventually and probably collapse into a BH once relativistic instability is triggered~\cite{Feng:2021rst}. It serves as the prototype of supermassive BHs from direct collapse of pristine gas~\cite{Begelman:2006db,Shang:2009ij,Agarwal:2012,Dunn:2018,Wise:2019} or dark matter halo with self-interaction~\cite{Balberg:2001qg,Balberg:2002ue,Pollack:2014rja,Shapiro:2018vju,Feng:2020kxv}.

As long as heat transport occurs, a self-gravitating monatomic fluid (or supermassive star) will relax and shrink automatically due to the negative specific heat of a gravitationally bound system~\cite{Balberg:2002ue,Feng:2020kxv}. 
During the gravothermal process in the hydrostatic limit, the thermal evolution timescale of the contraction is much larger its free-fall (dynamical) timescale $t_{\rm ff}\sim1/\sqrt{G_N\rho}$, where $\rho$ is the (mean) energy density of the fluid~\cite{Shapiro:1983du}. In this scenario, we can idealize the evolution process by a sequence of virialized quasiequilibria characterized by the mass and radius of the fluid sphere. In particular, the particles in the fluid will follow the same distribution function, albeit the dispersion varies during the process until the onset of relativistic instability. Moreover, we assume \emph{no} extra degrees of freedom, {e.g.}, nuclear reaction of our universe in (3+1) dimensions, will be ignited to halt the direct collapse into a BH.
We note that the final BH formation near the end of the gravothermal evolution requires dynamically evolving the fluid and the spacetime given initial data~\cite{Markovic:1999di,Goswami:2006ph,Lai:2007tj,Noble:2015anf}, which is beyond the scope of this study. 
Without a cosmological constant in (3+1), it has been shown that unstable static spherical Tolman-Oppenheimer-Volkoff solutions exist on saddle points that, when perturbed from their unstable equilibrium, will tend to either black hole formation or a perturbed stable solution~\cite{Noble:2015anf}; 
while with positive cosmological constant, the dynamical evolution of a homogeneous dust would drag the entire spacetime into a ``big crunch'' singularity if the fluid mass is sufficiently large~\cite{Markovic:1999di}.



The goal of this {paper} is to examine the \emph{sufficient} condition that can \emph{naturally} trigger the instability of a self-gravitating monatomic fluid in ($N$+1) dimensions, in particular, in the presence of cosmological constant. We adopt homogeneous solutions, which are adequate for the purpose.
In the end, we will briefly discuss the implications on the dimensionality of spacetime. The geometric unit $G_N=c=1$ is used, unless noted otherwise.

\paragraph*{Dynamical instability in ($N$+1) dimensions.---}
 The method exploited by Chandrasekhar~\cite{Chandrasekhar:1964zza} is to examine the  \emph{radial pulsation equation} of a perturbed fluid sphere of mass $\mathcal{M}$ within radius $R$:
\begin{equation}
\delta\ddot{R}+\omega^2 \delta R=0 
\quad{\rm with}\quad
\omega^2\propto \langle\gamma\rangle-\gamma_{\rm cr},
\end{equation}
where $\omega$ is oscillation frequency, the critical adiabatic index $\gamma_{\rm cr}$ depends on the given equilibrium configuration, and $\langle\gamma\rangle$ is the pressured-averaged adiabatic index of the fluid sphere. Thus the stability problem boils down to the Sturm-Liouville eigenvalue problem. The \emph{sufficient} condition for the fluid to become unstable is $\gamma_{\rm cr}>\langle\gamma\rangle$ such that $\omega^2<0$, implying the perturbation $\delta R\sim e^{i\omega t}$ would be an exponential growth.

The adiabatic index of a fluid,
\begin{equation}
\gamma=\left(\frac{\partial\ln p}{\partial\ln n}\right)_{s},
\end{equation}
is a \emph{stiffness/compressibility} parameter signifying how the fluid pressure $p$ responds to the adiabatic (${\rm d}s=0$) compression on number density $n$.
In particular, an ideal fluid parametrized by the $\gamma$-law form $p=K(mn)^\gamma$~\cite{Tooper:1965}, where $m$ is the particle's mass, satisfies the above definition
as long as $\gamma$ and $K$ are not explicit functions of $n$ under adiabatic perturbation. The first law of thermodynamics results in~\cite{Weinberg:1972kfs,Shapiro:1983du,Feng:2021acd}
$\gamma=1+p/(\rho-mn)$.
Given a distribution function  $f(\mathbf{x}, \mathbf{p})$ of monatomic particles with phase space measure ${\rm d}^N\mathbf{x}~{\rm d}^N\mathbf{p}$, the adiabatic index of the ideal fluid merely depends on its \emph{velocity dispersion} 
$v\equiv \sqrt{Np/\rho}<1$ and the \emph{degrees of freedom} $N$, specifically 
\begin{equation}
\gamma=1+\frac{1+\sqrt{1-v^2}}{N}
\end{equation}
ranges from  $1+2/N$ (nonrelativistic $v\rightarrow0$) to $1+1/N$ (ultrarelativistic $v\rightarrow1$)~\cite{Feng:2021acd}.

Considering NG in ($N$+1)-dimensional spacetime, one can derive
the \emph{critical adiabatic index}~\cite{Feng:2021acd}:
\begin{equation}
\gamma_{\rm cr (NG)}=2\left(1-\frac{1}{N}\right).
\end{equation}
In order to have a stable configuration, it is \emph{necessary} that $\langle\gamma\rangle>\gamma_{\rm cr (NG)}$. For ultrarelativistic (nonrelativistic) ideal fluids, this implies the spatial dimensions must be $N<3~(N<4)$ in order to have a stable sphere. From this viewpoint, the privilege of (3+1) dimensions is manifest because \emph{the fluid sphere is stable but not too stable.} However, in (2+1) dimensions the fluid disk is too stable because $\gamma_{\rm cr (NG)}=1<1.5~(2)=\gamma$ as always for an ultrarelativistic (nonrelativistic) fluid. 
Nevertheless, in the context of GR the pressure effect is crucial to destabilize the fluid disk. Besides, in order for the fluid to have equilibrium configurations and a BTZ solution, a \emph{negative cosmological constant} is required~\cite{Banados:1992wn}.

Thus we have to first examine the equilibrium configurations with cosmological constant $\lambda=\pm1/\ell^2$ in ($N$+1) dimensions. For homogeneous solutions, 
the critical adiabatic index\footnote{The exact expression is derived in Ref.\cite{Feng:2021acd} and see also Supplemental Material for a full GR expression.}
\begin{align}
\gamma_{\rm cr (GR)}=
\frac{\lambda R^2}{(N-2)\mathcal{M}/R^{N-2}-\lambda R^2}\nonumber\\
+\sum_{j, k=0,1,...}f_{jk}^{(N)}\left(\frac{\mathcal{M}}{R^{N-2}}\right)^j \left(\lambda R^2\right)^k ,
\end{align}
where the post-Newtonian coefficients $f_{jk}^{(N)}$ depend on the density distribution and spatial dimensions $N$, except $f_{00}^{(N)}=\gamma_{\rm cr (NG)}$;
and the \emph{stabilizer/destabilizer}:
\begin{equation}
\frac{\lambda R^2}{(N-2)\mathcal{M}/R^{N-2}-\lambda R^2}
\begin{cases}
~{\rm stabilizer~if~negative}\\
~{\rm destabilizer~if~positive}
\end{cases}
\end{equation}
characterizes the relative competition between compactness and background curvature. We note that its appearance is generically from GR as long as $\lambda$ is switched on, and \emph{cannot} be regarded as post-Newtonian correction. 

Qualitatively, the GR instability depends on the pressure effect of the fluid through 
$p/\rho\propto\mathcal{M}/R^{N-2}\equiv\mathcal{C}_N$, the \emph{compactness parameter} in ($N$+1) dimensions.
On the other hand, the stability of a fluid will also depend on the relative size of the fluid to the radius of curvature of the space, specifically, the curvature parameter $\lambda R^2$.
As was mentioned, in the context of NG, fluid spheres are genuinely subject to dynamical instability for $N>3$. Even worse, the corrections from GR deteriorate the situation, especially if $\lambda>0$. However, it is possible to have stable hydrostatic equilibrium if $\lambda<0$.
We also note that for $\lambda=0$ the post-Newtonian approximation  $f^{(3)}_{10}=19/21$ is exactly the result shown in Ref.~\cite{Chandrasekhar:1964zza}.

\paragraph*{Gravitationally bound systems.---}
In GR, the \emph{gravitational} mass of a fluid sphere $\mathcal{M}$ is the corresponding Schwarzschild mass ($N\geq3$) if it were to collapse into a BH. It includes the energy of self-gravity due to the curved spacetime, which is thus not conserved during the gravothermal evolution. By contrast, the rest mass $\mathcal{M}_{\rm rest}$ of the fluid is conserved (see Appendix~B for definition). It is the mass of total particles in the fluid when they are dispersed to infinity. Therefore, to form a gravitationally bound state the total internal energy must be 
\begin{equation}
\mathcal{M}-\mathcal{M}_{\rm rest}<0.
\end{equation}
Before we are able to examine the dynamical instability reasonably, it is necessary to see if the quasistatic equilibrium is gravitationally bound during the gravothermal evolution. The solutions can be categorized as stable or unstable \emph{only} if they are gravitationally bound. If the \emph{initial} configuration is a unbound state, $\mathcal{M}-\mathcal{M}_{\rm rest}>0$, 
dynamical evolution of the fluid and the spacetime is required to determine the final fate (BH or naked singularity)~\cite{Goswami:2006ph}, which is again beyond the scope of the paper.

\begin{figure}[htbp]
\centering
   \includegraphics[width=0.48\textwidth]{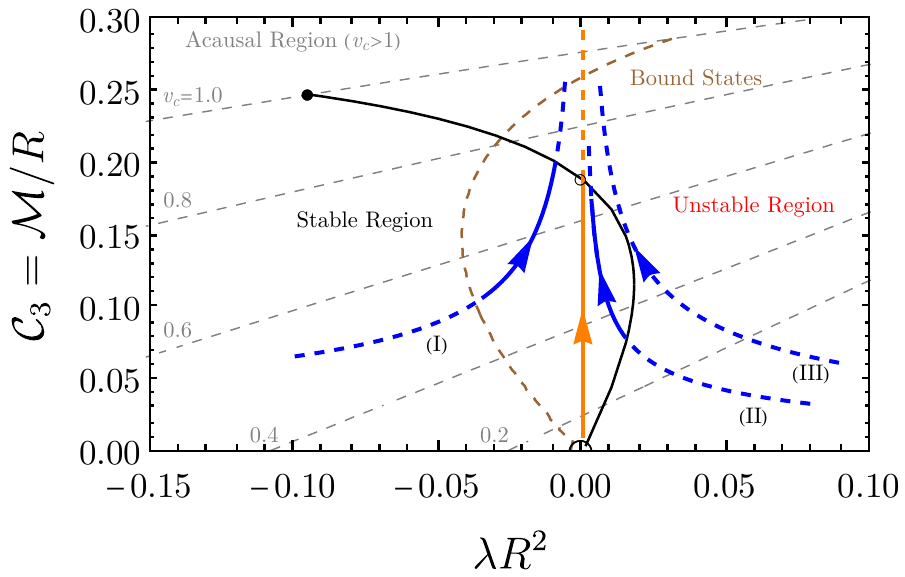}
   \caption{ $\mathcal{C}_3-\lambda R^2$ phase diagram of homogeneous fluid spheres in (3+1) dimensions. Bound states are to the right of the brown dashed line. The stable and unstable regions are separated by the marginal stable curve (black solid), and the black dot denotes the end point $(-0.0949, 0.248)$ at the causal limit. As the radius contracts with $\mathcal{M}_{\rm rest}={\rm const}$, the orange path follows $\lambda=0$, and the circle denotes  the critical point $(0, 0.189)$ of instability; the blue paths (I), (II), (III) follow $\mathcal{M}_{\rm rest}\sqrt{\abs{\lambda}}=0.02(\lambda<0), 0.01(\lambda>0)$, $0.02(\lambda>0)$, respectively.}
   \label{fig:n3phase}
\end{figure}

\begin{figure}[htbp]
\centering
   \includegraphics[width=0.48\textwidth]{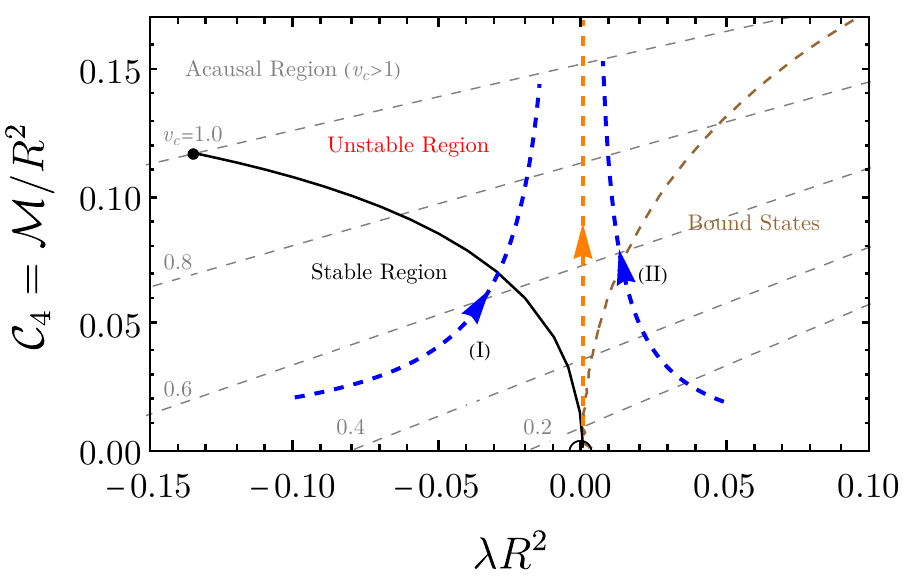}
   \caption{ $\mathcal{C}_4-\lambda R^2$ phase diagram of homogeneous fluid spheres in (4+1) dimensions. Bound states are to the right of the brown dashed line. The stable and unstable regions are separated by the marginal stable curve (black solid), and the black dot denotes the end point $(-0.135, 0.118)$ at the causal limit. As the radius contracts with $\mathcal{M}_{\rm rest}={\rm const}$, the orange path follows $\lambda=0$, which is unbound;  the blue paths (I), (II) follow $\mathcal{M}_{\rm rest}\abs{\lambda}=0.002(\lambda<0), 0.001(\lambda>0)$, respectively.}
   \label{fig:n4phase}
\end{figure}

\paragraph*{Fluid spheres in  (3+1) and higher dimensions.---} 
Assuming a fluid sphere in ($N$+1) dimensions is in hydrostatic quasiequilibrium, the fluid (rest) mass $\mathcal{M}_{\rm rest}={\rm const}$ during the gravothermal evolution. The radius will contract such that $\mathcal{C}_N$ increases gradually as more and more thermal energy dissipates until reaching the critical compactness as $\langle\gamma\rangle=\gamma_{{\rm cr}}$.
 Given $\lambda={\rm const}$, we can tell from the phase diagrams (Figs.~\ref{fig:n3phase}~and~\ref{fig:n4phase}) when the phase transition into BH could be triggered. For $N\geq3$ the evolution follows $\mathcal{M}_{\rm rest}\abs{\lambda}^{(N-2)/2}={\rm const}$, only those paths passing through the stable bound region will be in the ``long-lived'' stage of gravothermal evolution.
  
As we have already noted, a fluid sphere is genuinely unstable for $N\geq4$, and $\lambda>0$ just deteriorates the situation.
Remarkably, the privileged position of $N=3$ can be seen also from the fact that \emph{it is the unique dimensionality that allows stable hydrostatic equilibrium with positive cosmological constant.}
From Fig.~\ref{fig:n3phase}, we see  that for $\lambda>0$ stable hydrostatic equilibrium exists only between some upper bound and lower bound of compactness, and the stable region of compactness diminishes as $\lambda R^2$ increases; for $\lambda<0$, the stable region enlarges as $\abs{\lambda R^2}$ increases until the critical compactness $\mathcal{C}_3=0.248$ at causal limit $v(0)\equiv v_c=1$. 
However, bound states no longer exist well before this critical point.
The orange path ($\lambda=0$) will gravothermally transition from a stable region into an unstable one after passing critical compactness $\mathcal{C}_3=0.189$ and the collapse into BH might ensue. The blue path (I), which follows $\mathcal{M}_{\rm rest}\sqrt{\abs{\lambda}}=0.02(\lambda<0)$, will be gravothermally transitioning from the stable region into the unstable one after hitting the marginal stable curve if it starts from the region of bound states; while the blue path (II), which follows $\mathcal{M}_{\rm rest}\sqrt{\abs{\lambda}}=0.01(\lambda>0)$, starting from the unstable region could directly collapse into the stable region of gravothermal evolution until exceeding the upper critical compactness; however, if the mass is sufficiently heavy as blue (dashed) path (III) with $\mathcal{M}_{\rm rest}\sqrt{\abs{\lambda}}=0.02(\lambda>0)$, there is no long-lived gravothermal evolution of the fluid. 

On the other hand, in Fig.~\ref{fig:n4phase}, we see that for $N=4$ there is no stable hydrostatic equilibrium for $\lambda\geq0$. A stable region emerges if $\lambda<0$, and the critical compactness increases as $\abs{\lambda R^2}$ increases until $\mathcal{C}_4=0.118$ at $v_c=1$. Nevertheless, no bound state exists in the domain of $\lambda\leq0$. 
For instance, there is no stable bound state along the orange dashed path ($\lambda=0$). 
Although the blue dashed path (I), following $\mathcal{M}_{\rm rest}\abs{\lambda}=0.002 (\lambda<0)$, could transition from the stable region into an unstable one, it is by no means gravothermal as no static bound state is available along this path. Finally, the blue dashed path (II), which follows $\mathcal{M}_{\rm rest}\abs{\lambda}=0.001 (\lambda>0)$, always lies in the unstable region no matter if it starts from a bound or an unbound state. 
The phase diagrams are similar for $N>4$ but it becomes less compact on the marginal stable curves as $N$ increases. In Table~\ref{tab:endpoint}, we show the end points ($v_c=1$) of the marginal stable curves for $N=2,3,4,5,6$ and $7$. 

Remarkably, the region of bound states \emph{never} overlaps with the stable region for $N\geq4$.
Dynamically, if the fluid starts from any point on the dashed paths in Figs.~\ref{fig:n3phase}~and~\ref{fig:n4phase}, BH formation, dispersal of the fluid to infinity, or gravitationally bound and oscillatory states could be the possible outcome depending on the initial velocity perturbation and density~\cite{Noble:2015anf}, which deserves further investigation.

\setlength{\tabcolsep}{20pt} 
\renewcommand{\arraystretch}{1} 
\begin{table}
  \caption{End points of marginal stable curves for $N=2,3,4,5,6,$ and $7$ with $\lambda<0$ at causal limit $v_c=1$.}
 \resizebox{\columnwidth}{!}{%
 \begin{tabular}{c c c c}
  \hline\hline
  $N$  &  $\mathcal{C}_N$  & $\lambda R^2$ & $\langle\gamma\rangle=\gamma_{{\rm cr}}$ \\ [0.5ex] 
  \hline\hline
$2$ & $0.518001$ & $-0.060912$ & $1.81893$ \\
$3$ & $0.248179$ & $-0.094853$ & $1.56387$ \\
$4$ & $0.117505$ & $-0.134605$ & $1.43352$ \\
$5$ & $0.062846$ & $-0.151149$ & $1.35328$ \\
$6$ & $0.037099$ & $-0.154395$ & $1.29861$ \\
$7$ & $0.023595$ & $-0.151406$ & $1.25884$ \\
[0.5ex]
  \hline\hline
 \end{tabular}
 }
 \label{tab:endpoint}
\end{table}

\paragraph*{Fluid disks in (2+1) dimensions.---} 
By matching the junction conditions, the mass of the BTZ BH is related to the gravitational mass of the fluid disk by~\cite{Feng:2021acd}
\begin{equation}
\mathcal{M}_{\rm BTZ}=2\mathcal{M}-1,
\end{equation}
thus $\mathcal{M}>0.5$ is the threshold to have $\mathcal{M}_{\rm BTZ}>0$, the excited state, if collapse ensues.
Dynamical collapse into BTZ BHs and naked singularities has been shown possible from pressureless dust~\cite{Ross:1992ba}.
However, static stars of perfect fluid qualitatively differ in their behavior from static stars of dust~\cite{Giddings:1983es}; it is curious to see if the GR instability will be triggered in (2+1), especially under the influence of a negative cosmological constant.

Therefore we have to examine the critical adiabatic index for $N=2$:
\begin{equation}
\gamma_{\rm cr (GR)}=-1
+\sum_{j, k=0,1,...}f_{jk}^{(2)}\mathcal{M}^j \left(\lambda R^2\right)^k
\end{equation}
starts from $-1+f_{00}^{(2)}=-1+\gamma_{\rm cr (NG)}=0$ with ``post-Newtonian'' corrections, thus
\emph{the Einsteinian stars are much stabler than Newtonian stars in (2+1).}
We note that for $N=2$ the ``compactness'' parameter reduces to $\mathcal{M}$, the gravitational mass of the disk itself. That manifests the reason why there is no Buchdahl-like bound in (2+1) dimensions~\cite{Cruz:1994ar}. 
Furthermore, this also implies that \emph{a self-gravitating disk cannot gravothermally evolve to a singular state on its own} because the compactness $\mathcal{M}$ always \emph{decreases} with $\mathcal{M}_{\rm rest}={\rm const}$ due to the gravothermal dissipation. However, it can become unstable by external agents, such as compression by external force while adding mass to keep $\mathcal{M}$ large~\cite{Feng:2021acd}.

\begin{figure}[htbp]
\centering
   \includegraphics[width=0.48\textwidth]{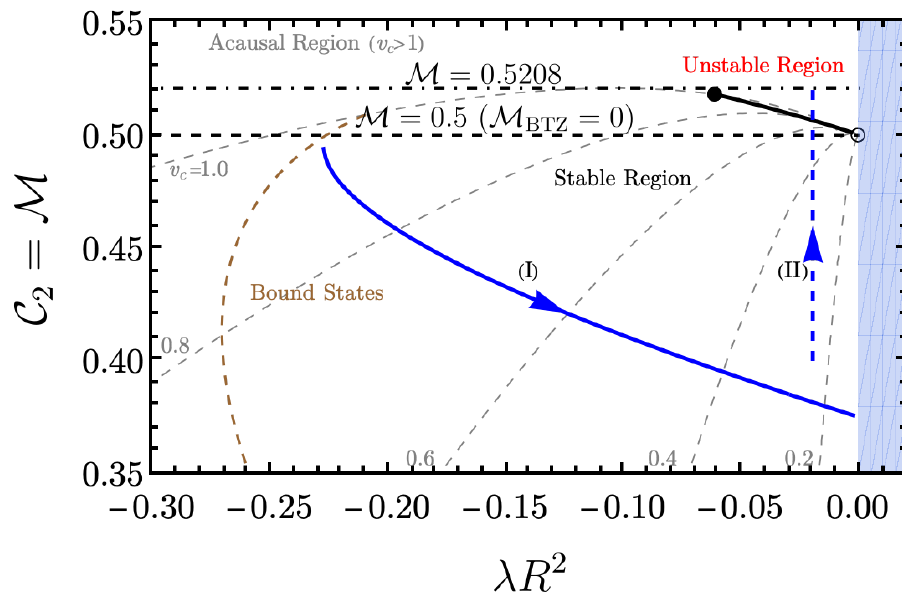}
   \caption{ $\mathcal{C}_2-\lambda R^2$ phase diagram of homogeneous fluid disks in (2+1) dimensions. Bound states are to the right of the brown dashed line. The stable and unstable regions are separated by the marginal stable curve (solid black), and the black dot denotes the end point $(-0.0609, 0.518)$ at the causal limit. The shaded region ($\lambda>0$) is forbidden to have hydrostatic equilibrium. The circle denotes $\mathcal{M}=0.5$ exactly at $\lambda=0$, which is independent of central velocity dispersion~\cite{Giddings:1983es}. The upper bound of fluid mass is $\mathcal{M}=0.5208$ at the causal limit. The region under $\mathcal{M}=0.5$ is stable, which means that no homogeneous fluid disk can trigger the instability and collapse into a naked singularity. Path (I) follows $\mathcal{M}_{\rm rest}=0.5$ under gravothermal evolution; (II)  follows $\lambda R^2=-0.02$ by adding mass. Only path (II) could transition into a BTZ BH under the causal limit.}
   \label{fig:n2phase}
\end{figure}

 To illustrate, we see from Fig.~\ref{fig:n2phase} that as the negative cosmological constant is switched on, there is no unstable configuration for $\gamma_{\rm cr}$ to cross $\langle\gamma\rangle$ of the fluid disk as the fluid mass grows from $\mathcal{M}=0$ to $0.5$ (the would-be $\mathcal{M}_{\rm BTZ}=-1$ to $\mathcal{M}_{\rm BTZ}=0$), which means there is no instability for a fluid disk to collapse into a naked singularity from an ideal fluid within causal region $v_c\leq1$. 
Therefore, along the solid blue path (I), $\mathcal{M}_{\rm rest}=0.5$ in Fig.~\ref{fig:n2phase},  the compactness decreases from $\mathcal{M}=0.495$ to $0.375$ during the gravothermal shrinking, it never meets the instability. That is to say, although dissipation or thermal radiation can make the disk shrink naturally given $\lambda={\rm const}$, it never drives the fluid into an unstable state. 
Nevertheless, in the range $-0.061\lesssim\lambda R^2<0$ a BTZ BH could emerge from a collapsing fluid of $0.5<\mathcal{M}\lesssim0.518$ without violating causality. For example, under the background $\lambda={\rm const}$ this can be achieved by ``adding more mass'' to the fluid disk, while the radius remains fixed, as shown by the dashed blue path (II): $\lambda R^2=-0.02$ in Fig.~\ref{fig:n2phase}.

\paragraph*{Discussions and implications.---} 
In the context of gravothermal evolution, we have examined the dynamical instability of a self-gravitating fluid sphere in ($N$+1)-dimensional spacetime by adopting homogeneous fluid solutions. Although the critical $\mathcal{C}_N$ may vary quantitatively depending on the density distribution, the main conclusion generally holds as it is based on the three assumptions made implicitly~\cite{Feng:2021acd}:
(i) The monatomic fluids obey the first law of thermodynamics, and the pressure is isotropic due to equipartition theorem. (ii) The particles composing the fluid follow the mass-energy dispersion relation. (iii) Gravity is governed by Einstein field equations in ($N$+1) dimensions.

From the dynamical instability viewpoint, we can reexamine why (3+1) is privileged rather than why it must be (3+1). 
If BH is the pathway to generate a baby universe~\cite{Pathria:1972,Frolov:1988vj,Frolov:1989pf,Easson:2001qf,Poplawski:2010kb,Brandenberger:2021ken}, the collapsing matter squeezing into a ($N$+1)-dimensional BH near the classical singularity would result in a new-born universe of arbitrary dimensions. 
If the spacetime dimensionality reshuffling~\cite{Tegmark:1997jg} is a random process in the reign of quantum gravity near singularity, it repeats this process again and again until the new-born universe is just (3+1)-dimensional, in which the fluid sphere is stable but not too stable. As a self-gravitating fluid sphere (or a star) in $N\geq4$ is genuinely subject to dynamical instability, and could transition into a BH automatically without undergoing the stage of long-lived gravothermal evolution. 
Although a fluid star can be stabilized by introducing a negative cosmological constant, no gravitationally bound state of a monatomic fluid could exist in this region.

Remarkably, (3+1) is the unique dimensionality that allows stable hydrostatic equilibrium with positive cosmological constant. Given the cosmological constant observed~\cite{Prat:2021xlz} (or $\ell\sim10^{61}\ell_{\rm Pl}$, where $\ell_{\rm Pl}$ is the Planck length), the  mass of a virialized stellar object, {e.g.}, dark matter halo, must be $\mathcal{M}\ll0.02\ell c^2/G_3\sim10^{21}~M_\odot$ (see Fig.~\ref{fig:n3phase}) in order to avoid the dynamical instability and a possible BH formation from its direct collapse. 
However, a big crunch singularity would form if $\mathcal{M}>(1/3\sqrt{3})\ell c^2/G_3\sim10^{22}~M_\odot$~\cite{Markovic:1999di}.
On the other hand, (2+1)-dimensional gravity is bizarre. The fluid disk cannot gravothermally evolve into a singular state through dissipation, but a BTZ hole could emerge from a collapsing fluid disk with external agents, while a naked singularity cannot emerge from a fluid disk, which supports the cosmic censorship conjecture~\cite{Penrose:1969pc}.

\section*{Acknowledgements}
The author acknowledges the Institute of Physics, Academia Sinica, where part of the work was completed, for their hospitality. The author is also grateful for helpful correspondence with Steve Carlip and Stanley Deser. In particular, the author thanks the anonymous referee for useful comments, which largely improve the paper. This work is supported in part by the U.S. Department of Energy under grant No. DE-SC0008541.

\bibliographystyle{JHEP} 
\bibliography{dibh}

\onecolumngrid
\appendix
\section*{Supplemental Material}
\section{Critical adiabatic index in ($N$+1)-dimensional spacetime}
In ($N$+1) dimensions, ${\rm d}s^2=-e^{2\Phi(t,r)}{\rm d}t^2+e^{2\Lambda(t,r)}{\rm d}r^2+r^2{\rm d}\Omega_{N-1}^2$ one can show that
\begin{align}
\nonumber
\gamma_{\rm cr (GR)}=
&2\left(1-\frac{1}{N}\right)
+\frac{\int e^{3\Phi+\Lambda}[8(N-1)p+(N-2)(e^{2\Lambda}-1)(\rho+p)](N-2)(e^{2\Lambda}-1)r^{N-1}{\rm d}r}{4N^2\int e^{3\Phi+\Lambda}pr^{N-1}{\rm d}r}\\
\nonumber
&+\frac{\kappa_N\int e^{3(\Phi+\Lambda)}\left\{[4(N-1)p+(N-2)(e^{2\Lambda}+1)(\rho+p)](p+p_\lambda)-2(N-1)p_\lambda(\rho+p)\right\} r^{N+1}{\rm d}r}{N^2(N-1)\int e^{3\Phi+\Lambda}pr^{N-1}{\rm d}r}\\
&+\frac{\kappa_{N}^2\int e^{3\Phi+5\Lambda}(\rho+p)(p+p_\lambda)^2r^{N+3}{\rm d}r}{N^2(N-1)^2\int e^{3\Phi+\Lambda}pr^{N-1}{\rm d}r},
\end{align}
where $\kappa_{N}=(N-1)\omega_{N}$ is the (Einstein's) gravitational constant in $N$-dimensional space ($G_N=c=1$) with
\begin{equation}
\omega_N=\frac{2\pi^{N/2}}{\Gamma(N/2)}
\end{equation}
being the area of the unit sphere in $N$-dimensional space; and
\begin{equation}
\langle\gamma\rangle\equiv\frac{\int e^{3\Phi+\Lambda}\gamma pr^{N-1}{\rm d}r}{\int e^{3\Phi+\Lambda}pr^{N-1}{\rm d}r}
\end{equation}
the ``effective'' (pressure-averaged) adiabatic index of the fluid sphere.

\section{Gravitational and rest masses of fluid sphere in ($N$+1)-dimensional spacetime}
Due to the radiation of internal (thermal) energy during gravothermal evolution, the \emph{gravitational mass} of the fluid sphere
\begin{equation}
\mathcal{M}\equiv\omega_N\int \rho(r)r^{N-1}dr
\end{equation}
is \emph{not conserved}; whereas the \emph{rest mass} of the sphere
\begin{equation}
\mathcal{M}_{\rm rest}\equiv\omega_N\int mn(r)e^{\Lambda(r)}r^{N-1}dr
\end{equation}
is \emph{conserved}, where the rest mass density $mn(r)=\rho(r)\sqrt{1-v^2(r)}$ with $N$-dimensional velocity dispersion $v=\sqrt{Np/\rho}$. The total internal energy, including the gravitational potential, must be $\mathcal{M}-\mathcal{M}_{\rm rest}<0$ for a gravitationally bound system.

\section{Homogeneous solution in ($N$+1)-dimensional spacetime}
The profile of the pressure of an homogeneous density ($\rho={\rm const}$) fluid sphere of radius $R$ is given by
\begin{equation}
\frac{p(r)}{\rho}=\frac{ [N\rho-2(\rho+\rho_\lambda)]\left[ e^{-\Lambda(r)} - e^{-\Lambda(R)}\right] }{ N\rho~e^{-\Lambda(R)} - [N\rho-2(\rho+\rho_\lambda)]e^{-\Lambda(r)} },
\end{equation}
and the metric potential
\begin{equation}
e^{\Phi(r)}=\frac{ N\rho~e^{-\Lambda(R)} - [N\rho-2(\rho+\rho_\lambda)]e^{-\Lambda(r)} }{ 2(\rho+\rho_\lambda) },
\end{equation}
where
\begin{equation}
e^{-\Lambda(r)}=\sqrt{1-\frac{2\kappa_N(\rho+\rho_\lambda)}{N(N-1)}r^2}.
\end{equation}
In terms of gravitational mass $\mathcal{M}$, we can express the fluid density
\begin{equation}
\rho=\frac{N(N-1)\mathcal{M}}{\kappa_N R^N}
\end{equation}
and the cosmological constant
\begin{equation}
\rho_\lambda=\frac{N(N-1)}{2\kappa_N}\lambda=-p_\lambda,
\end{equation}
where $\lambda=\pm1/\ell^2$, the ``$+/-$'' sign corresponds to the positive/negative cosmological constant (scalar curvature). Incidentally, demanding $p(0)<\infty$ and $e^{\Phi(0)}>0$ leads to
the Buchdahl bound in ($N+1$)-dimension with cosmological constant $\lambda$:
\begin{equation}
\frac{N-1}{N^2}\left(1-\sqrt{1-\frac{N^2a}{(N-1)^2}}\right)
<
\mathcal{C}_N
<
\frac{N-1}{N^2}\left(1+\sqrt{1-\frac{N^2a}{(N-1)^2}}\right),
\end{equation}
where the compactness $\mathcal{C}_N=\mathcal{M}/R^{N-2}$ and the curvature parameter $a=\lambda R^2$.
Furthermore, the solution can be parametrized in terms of $\mathcal{C}_N$ and $a$:
\begin{equation}
\frac{p(x)}{\rho}
=\frac{\left[\left(N-2\right)\mathcal{C}_N-a\right]\left[\sqrt{1-\left(2\mathcal{C}_N+a\right)x^2}-\sqrt{1-\left(2\mathcal{C}_N+a\right)}\right]}{N\mathcal{C}_N\sqrt{1-\left(2\mathcal{C}_N+a\right)}-\left[\left(N-2\right)\mathcal{C}_N-a\right]\sqrt{1-\left(2\mathcal{C}_N+a\right)x^2}}
\equiv y\left(\mathcal{C}_N, a; x\right)
\end{equation}
and
\begin{equation}
e^{\Phi(x)}=\frac{N\mathcal{C}_N\sqrt{1-\left(2\mathcal{C}_N+a\right)}-\left[\left(N-2\right)\mathcal{C}_N-a\right]\sqrt{1-\left(2\mathcal{C}_N+a\right)x^2}}{2\mathcal{C}_N+a},
\end{equation}
where $x=r/R$ is the normalized radius, and we have used
\begin{equation}
e^{-\Lambda(x)}=\sqrt{1-\left(2\mathcal{C}_N+a\right)x^2}.
\end{equation}
The parametrization makes sense only when $\rho\neq0$ or $\mathcal{C}_N\neq0$.
With all of the above, the rest mass compactness (B2) turns out to be:
\begin{equation}
\mathcal{C}_{{\rm rest},N}\equiv\frac{\mathcal{M}_{\rm rest}}{R^{N-2}}
=N\mathcal{C}_N\int_0^1\sqrt{1-Ny}~e^\Lambda x^{N-1}dx.
\end{equation}
Moreover, the critical adiabatic index for homogeneous sphere can be expressed as:
\begin{align}
\nonumber
\gamma_{\rm cr (GR)}\equiv
&2\left(1-\frac{1}{N}\right)
+\frac{\int_0^1 e^{3\Phi+\Lambda}[8(N-1)y+(N-2)(e^{2\Lambda}-1)(1+y)](N-2)(e^{2\Lambda}-1)x^{N-1}{\rm d}x}{4N^2\int_0^1 e^{3\Phi+\Lambda}y x^{N-1}{\rm d}x}\\
\nonumber
&+\frac{\mathcal{C}_N\int_0^1 e^{3(\Phi+\Lambda)}\left\{[4(N-1)y+(N-2)(e^{2\Lambda}+1)(1+y)](y-a/2\mathcal{C}_N)+(N-1)(a/\mathcal{C}_N)(1+y)\right\} x^{N+1}{\rm d}x}{N\int_0^1 e^{3\Phi+\Lambda}yx^{N-1}{\rm d}x}\\
&+\frac{\mathcal{C}_{N}^2\int_0^1 e^{3\Phi+5\Lambda}(1+y)(y-a/2\mathcal{C}_N)^2x^{N+3}{\rm d}x}{\int_0^1 e^{3\Phi+\Lambda}yx^{N-1}{\rm d}x}.
\end{align}
In the ``post-Newtonian'' approximations ($\mathcal{C}_N\ll1$ and $a\ll1$), we have
\begin{equation}
\gamma_{\rm cr (GR)}=
\frac{a}{\left(N-2\right)\mathcal{C}_N-a}
+\sum_{j, k=0,1,...}f_{jk}^{(N)}\mathcal{C}_N^j a^k,
\end{equation}
where $f_{00}^{(N)}=\gamma_{\rm cr (NG)}=2\left(1-1/N\right)$, and the \emph{stabilizer/destabilizer} characterizing the \emph{competition} between the compactness and the background curvature (cosmological constant) can be expanded as
\begin{equation}
\frac{a}{\left(N-2\right)\mathcal{C}_N-a}
=
\begin{cases}
&\sum_{n=1}^\infty\left(a/\left(N-2\right)\mathcal{C}_N\right)^n 
\quad{\rm if}~\abs{a}<\left(N-2\right)\mathcal{C}_N \\
&-1-\sum_{n=1}^\infty\left[\left(N-2\right)\mathcal{C}_N/a\right]^n
 \quad{\rm if}~\abs{a}>\left(N-2\right)\mathcal{C}_N.
\end{cases}
\end{equation}
The post-Newtonian expansion terms in different $N=2,3,4,5,6,7$ are:
\begin{align}
\sum_{j, k=0,1,...}f_{jk}^{(2)}\mathcal{C}_2^j a^k=1
&+\left(-\frac{1}{3}a-\frac{11}{72}a^2-\frac{37}{432}a^3+...\right)
+\left(-\frac{3}{4}a-\frac{3}{4}a^2-\frac{125}{192}a^3+...\right)\mathcal{C}_2 
\nonumber\\
&+\left(-\frac{19}{12}a-\frac{733}{288}a^2-\frac{439}{144}a^3+...\right)\mathcal{C}_2^2
+\left(-\frac{157}{48}a-\frac{133}{18}a^2-\frac{821}{72}a^3+...\right)\mathcal{C}_2^3
+\mathcal{O}\left(\mathcal{C}_2^4\right),
\\
\sum_{j, k=0,1,...}f_{jk}^{(3)}\mathcal{C}_3^j a^k=\frac{4}{3}
&+\left(-\frac{1}{3}a-\frac{11}{63}a^2-\frac{509}{4851}a^3+...\right)
+\left(\frac{19}{21}-\frac{197}{441}a-\frac{8303}{11319}a^2-\frac{6250253}{9270261}a^3+...\right)\mathcal{C}_3 
\nonumber\\
&+\left(\frac{850}{441}-\frac{334}{3773}a-\frac{536678}{280917}a^2-\frac{156471878}{64891827}a^3+...\right)\mathcal{C}_3^2 
\nonumber\\
&+\left(\frac{12994}{3087}-\frac{19468780}{9270261}a-\frac{13326694}{4991679}a^2-\frac{154134740724}{254830204629}a^3+...\right)\mathcal{C}_3^3
+\mathcal{O}\left(\mathcal{C}_3^4\right),
\\
\sum_{j, k=0,1,...}f_{jk}^{(4)}\mathcal{C}_4^j a^k=\frac{3}{2}
&+\left(-\frac{5}{16}a-\frac{23}{128}a^2-\frac{117}{1024}a^3+...\right)
+\left(\frac{13}{8}-\frac{1}{20}a-\frac{1489}{2560}a^2-\frac{29271}{51200}a^3+...\right)\mathcal{C}_4
\nonumber\\
&+\left(\frac{627}{160}+\frac{2681}{1280}a-\frac{749}{3200}a^2-\frac{169853}{204800}a^3+...\right)\mathcal{C}_4^2
\nonumber\\
&+\left(\frac{6081}{640}+\frac{139943}{12800}a+\frac{1004709}{102400}a^2+\frac{262717}{51200}a^3+...\right)\mathcal{C}_4^3
+\mathcal{O}\left(\mathcal{C}_4^4\right),
\\
\sum_{j, k=0,1,...}f_{jk}^{(5)}\mathcal{C}_5^j a^k=\frac{8}{5}
&+\left(-\frac{13}{45}a-\frac{793}{4455}a^2-\frac{4751}{40095}a^3+...\right)
+\left(\frac{11}{5}-\frac{59}{165}a-\frac{21007}{57915}a^2-\frac{2272373}{5733585}a^3+...\right)\mathcal{C}_5
\nonumber\\
&+\left(\frac{318}{55}+\frac{87902}{19305}a+\frac{313834}{147015}a^2+\frac{26267674}{1949189}a^3+...\right)\mathcal{C}_5^2
\nonumber\\
&+\left(\frac{32342}{2145}+\frac{1534568}{70785}a+\frac{63003670}{2166021}a^2+\frac{3294202586032}{158897134539}a^3+...\right)\mathcal{C}_5^3
+\mathcal{O}\left(\mathcal{C}_5^4\right),
\\
\sum_{j, k=0,1,...}f_{jk}^{(6)}\mathcal{C}_6^j a^k=\frac{5}{3}
&+\left(-\frac{4}{15}a-\frac{13}{75}a^2-\frac{629}{5250}a^3+...\right)
+\left(\frac{8}{3}+\frac{56}{75}a-\frac{99}{875}a^2-\frac{149819}{840000}a^3+...\right)\mathcal{C}_6
\nonumber\\
&+\left(\frac{112}{15}+\frac{6184}{875}a+\frac{6159}{1250}a^2+\frac{8108851}{21000000}a^3+...\right)\mathcal{C}_6^2
\nonumber\\
&+\left(\frac{10784}{525}+\frac{1745323}{52500}a+\frac{4713903}{87500}a^2+\frac{1155443729}{29400000}a^3+...\right)\mathcal{C}_6^3
+\mathcal{O}\left(\mathcal{C}_6^4\right),
\\
\sum_{j, k=0,1,...}f_{jk}^{(7)}\mathcal{C}_7^j a^k=\frac{13}{7}
&+\left(-\frac{19}{77}a-\frac{1843}{11011}a^2-\frac{14459}{121121}a^3+...\right)
+\left(\frac{235}{77}+\frac{12193}{11011}a+\frac{17791}{121121}a^2+\frac{18543457}{294445151}a^3+...\right)\mathcal{C}_7
\nonumber\\
&+\left(\frac{98790}{11011}+\frac{1153426}{121121}a+\frac{2342609814}{294445151}a^2+\frac{57331873990}{8791290937}a^3+...\right)\mathcal{C}_7^2
\nonumber\\
&+\left(\frac{3114030}{121121}+\frac{5442308}{121121}a+\frac{730021819446}{8791290937}a^2+\frac{523891374973556}{8800082227937}a^3+...\right)\mathcal{C}_7^3
+\mathcal{O}\left(\mathcal{C}_7^4\right).
\end{align}

\end{document}